\documentclass[aps,a4paper,twoside,twocolumn,secnumarabic,balancelastpage,amsmath, amssymb,superscriptaddress]{revtex4}

\usepackage{graphics}      
\usepackage[pdftex]{graphicx}      
\usepackage{longtable}     
\usepackage{bm}            
\usepackage{bbold}
\usepackage{verbatim}
\usepackage{mathtools}
\usepackage{adjustbox}
\usepackage[colorlinks=true]{hyperref}

\usepackage{dsfont}

\usepackage{stmaryrd}

\begin{document}

\title{Optical conductivity and resistivity in a four-band model for ZrTe$_5$ from \emph{ab-initio} calculations}

\date{\today}

\author{Corentin Morice}
\affiliation{Institute for Theoretical Physics Amsterdam and Delta Institute for Theoretical Physics, University of Amsterdam, 1098 XH Amsterdam, The Netherlands}
\author{Elias Lettl}
\affiliation{Center for Electronic Correlations and Magnetism, Experimental Physics VI, Institute of Physics, University of Augsburg, 86135 Augsburg, Germany}
\author{Thilo Kopp}
\affiliation{Center for Electronic Correlations and Magnetism, Experimental Physics VI, Institute of Physics, University of Augsburg, 86135 Augsburg, Germany}
\author{Arno P. Kampf}
\affiliation{Center for Electronic Correlations and Magnetism, Theoretical Physics III, Institute of Physics, University of Augsburg, 86135 Augsburg, Germany}

\begin{abstract}
ZrTe$_5$ is considered a potential candidate for either a Dirac semimetal or a topological insulator in close proximity to a topological phase transition. Recent optical conductivity results motivated a two-band model with a conical dispersion in 2D, in contrast to density functional theory calculations. Here, we reconcile the two by deriving a four-band model for ZrTe$_5$ using $\textbf{k} \cdot \textbf{p}$ theory, and fitting its parameters to the \emph{ab-initio} band structure. The optical conductivity with an adjusted electronic structure matches the key features of experimental data. The chemical potential varies strongly with temperature, to the point that it may cross the gap entirely between zero and room temperature. The temperature-dependent resistivity displays a broad peak, and confirms theoretically the conclusions of recent experiments attributing the origin of the resistivity peak to the large shift of the chemical potential with temperature.
\end{abstract}

\maketitle

\section{Introduction}

ZrTe$_5$ is currently attracting renewed attention from two different perspectives. The first was sparked when density functional theory (DFT) calculations claimed this material to be a strong topological insulator (STI) close to a topological phase transition towards a weak topological insulator (WTI) \cite{Weng2014}. Further computational work argued that this transition could be achieved via volume expansion \cite{Fan2017, Monserrat2019}. Experimentally, the situation is debated: a combination of scanning-tunnelling microscopy (STM), angle-resolved photoemission spectroscopy (ARPES), and \emph{ab-initio} calculations determined a gapped bulk electronic structure and topological states at the step edges, and ZrTe$_5$ was concluded to be WTI \cite{Li2016, Wu2016, Moreschini2016, Xiong2017}. Yet, this result was in contradiction with other work performed at the same time using the same experimental probes, which determined that it is an STI \cite{Manzoni2016}.

In the second perspective, ZrTe$_5$ is claimed to be a Dirac semimetal following the measurement of the chiral magnetic effect \cite{Li2016a}, as well as linear optical conductivity at low frequency \cite{Chen2015}. Reports of Shubnikov--de Haas (SdH) oscillations with a non-trivial Berry phase strengthened this claim, and unveiled a strongly anisotropic electronic structure \cite{Yuan2016, Zheng2016, Qiu2016}. The Landau level spectrum showed Dirac semimetal features, in addition to band inversion \cite{Chen2015a, Chen2017, Jiang2017}. Finally, a giant planar Hall resistivity, vanishing with the thickness of the sample, was observed and also interpreted as a signature of a Dirac semimetal \cite{Li2018}.

ZrTe$_5$ has been studied since the 1980s, before the rising interest in topological properties, owing to a prominent peak in the temperature dependent resistivity, in relation to a large thermoelectric power \cite{Jones1982, Skelton1982, Tritt1999}. ARPES measurements linked this peak to a crossover from $p$-type to $n$-type carriers when decreasing temperature \cite{McIlroy2004, Manzoni2015, Xiong2017, Zhang2017}, which concurred with a sign change of the Hall number \cite{Li2018, Martino2019}. The temperature of the resistivity maximum varied with sample thickness and doping \cite{Lu2017, Tang2018, Li2019}. Recently, this resistivity peak was also related to the potential topological properties of ZrTe$_5$ and interpreted as a signature for a transition upon cooling \cite{Nair2018} from a WTI to an STI state \cite{Xu2018} or the reverse \cite{Tian2019}, as well as a signature of Dirac polarons \cite{Fu2020}.

Differences between synthesis methods were shown to strongly influence the electronic properties, in particular for samples with $n$-doping in the case of chemical vapor transport growth, claimed to be responsible for the resistivity peak, as supported by \emph{ab-initio} calculations \cite{Shahi2018}.

Recently, a tentative minimal model for ZrTe$_5$ was put forward, based on optical conductivity measurements \cite{Martino2019, Wang2020, Rukelj2020}. In contrast to a 3D Dirac semimetal model, it features a linear dispersion in two directions and parabolic dispersion in the orthogonal direction. This important step attempts to describe theoretically experimental findings which have challenged theory until now. However, it falls short of completely uniting the field by also linking these experiments to the DFT results which initially sparked the interest for this material. In particular, this minimal model has a parabolic dispersion along the stacking direction, unlike the band structure calculated using DFT which features a double well structure. Additionally, it neglects the spin degree of freedom by considering only two bands, while DFT yields four bands, which are spin degenerate.

Here, we bridge this gap by starting from \emph{ab-initio} calculations and fitting an alternative minimal model, obtained from $\textbf{k}\cdot\textbf{p}$ theory, to the calculated bands. We adjust the electronic structure by modifying one parameter of the model, in particular changing the energy gap, known to be given inaccurately by DFT, to its experimental value. We calculate the optical conductivity, the temperature-dependent chemical potential and the resistivity, and find a frequency dependence of the optical conductivity matching experiments, without a parabolic dispersion along the stacking direction. The chemical potential strongly shifts at low temperature and thereby leads to a peak in the temperature-dependent resistivity, in agreement with ARPES and transport results.

\section{\emph{Ab-initio} calculations}
\label{Ab-initio calculations}

\begin{figure}
\centering
\begin{tabular}{cr}
a) & \raisebox{-0.95\height}{\includegraphics[width=8cm]{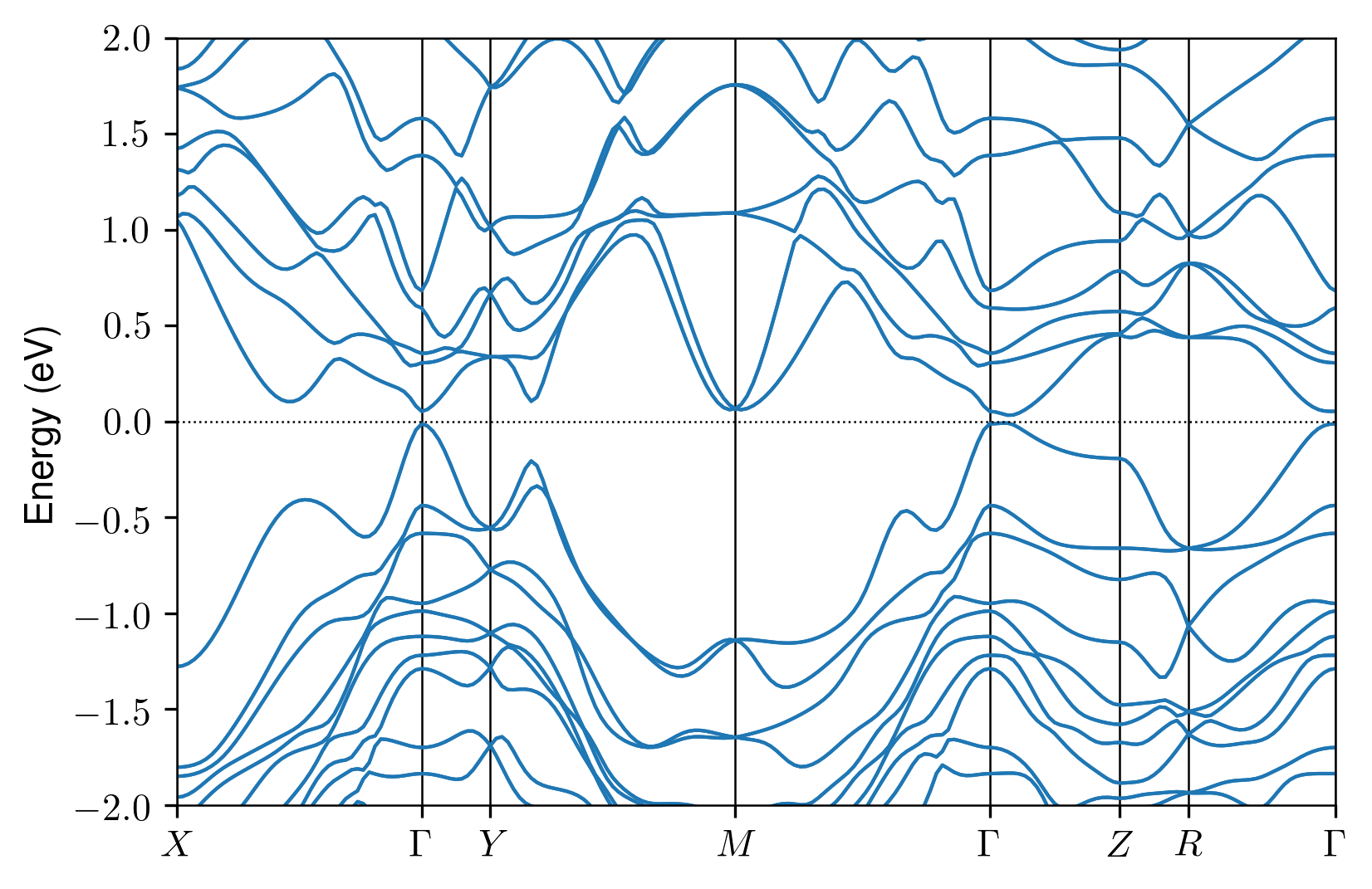}}
\\
b) & \raisebox{-0.95\height}{\includegraphics[width=8cm]{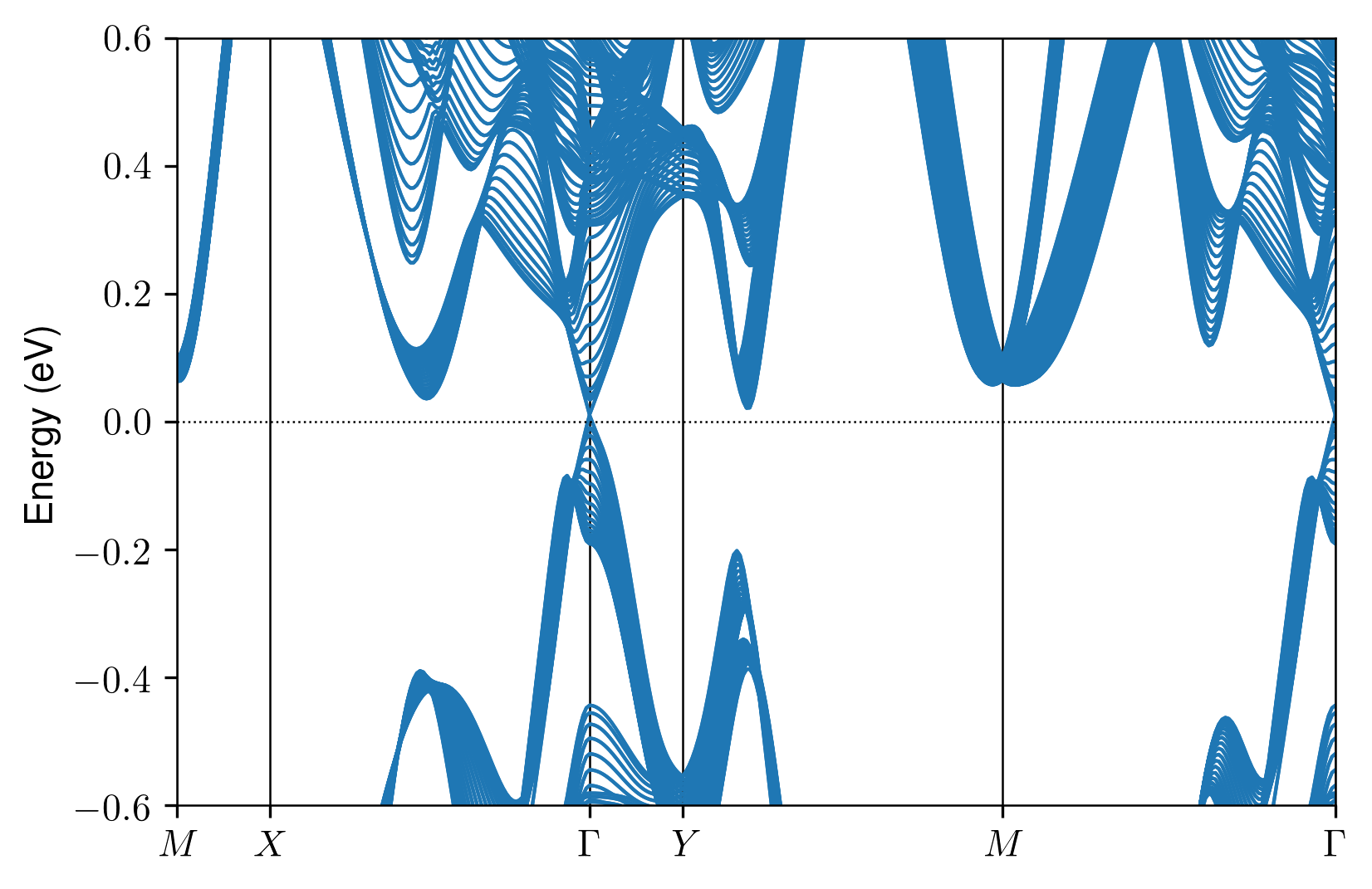}}
\end{tabular}
\caption{(a) Electronic structure of ZrTe$_5$ calculated using Elk. (b) Electronic structure of a slab of ZrTe$_5$ perpendicular to $b$. It displays surface states at $\Gamma$, consistent with the calculated topological invariants $(1,010)$. The $Z$ and $R$ points are located at finite $k_b$ and therefore do not exist in the Brillouin zone for the slab.}
\label{fig:DFT}
\end{figure}

We calculated the electronic structure using fully relativistic projector augmented-wave pseudopotentials and a plane-wave basis set as implemented in Quantum Espresso, and the all-electron full-potential linearized augmented plane-wave code Elk. We employed the generalised gradient approximation in the shape of the Perdew, Burke, and Ernzerhof functional \cite{Perdew1996}, and included spin-orbit coupling. We used the experimentally-measured crystal structure with $a = 3.9797$ \AA, $b=7.5036$ \AA\ and $c=13.676$ \AA\ \cite{Fjellvag1986}. ZrTe$_5$ is composed of tubes of Te atoms surrounding chains of Zr atoms. These tubes are linked together to form two-dimensional layers in the $a$-$c$ plane, which are coupled by van der Waals interactions along $b$. The calculations were performed with $16 \times 8 \times 4$ \textbf{k}-points in the primitive unit cell and a plane-wave energy cutoff of 70 Ry in Quantum Espresso \cite{Giannozzi2009}, and a predefined high-quality set of parameters in Elk \cite{Elk}. All the parameters were tested for convergence. The band structures calculated by the two methods are in close agreement. We constructed a tight-binding model by projecting the electronic structure obtained with Quantum Espresso on maximally localised Wannier orbitals using Wannier90 \cite{Wannier90}. The Zr $d$ orbitals and Te $p$ orbitals were included in the model. We then used this tight-binding model to calculate topological indices and slab band structures using Wannier Tools \cite{Wu2017}.

The band structure obtained in Elk is represented in Fig.\ \ref{fig:DFT}a. It features states close to the Fermi level at $\Gamma$, which get closer together on a short segment close to $\Gamma$ in the $\Gamma-Z$ direction before growing further apart. At $\Gamma$, the band gap is 65\ meV, while the minimum gap along $\Gamma-Z$ is 41\ meV. Even though Zr atoms have a partially filled $d$-shell and on-site correlations were not included in the calculations, the electronic structure exhibits a gap. This indicates that the physics of ZrTe$_5$ close to the Fermi level is not dictated by strong correlations. We calculated the 3D topological invariants $(\nu_0,\nu_1 \nu_2 \nu_3)$ to be $(1,010)$, and the band structure calculation for a slab perpendicular to the crystallographic $b$ direction (Fig.\ \ref{fig:DFT}b) gives surface states close to $\Gamma$, as expected for these invariants. These results are consistent with recent \emph{ab-initio} studies of ZrTe$_5$ \cite{Weng2014, Manzoni2016, Fan2017}.

\section{Minimal model}

We want to address the nature of the electronic structure of ZrTe$_5$ close to the Fermi level. DFT is instrumental in doing so, since it fits ARPES results very well, except for the magnitude of the gap \cite{Wu2016, Martino2019}. Many-body calculations, such as the GW approximation \cite{Hedin1969}, and hybrid exchange-correlation functionals, such as B3LYP \cite{Becke1993} and HSE \cite{Heyd2003}, significantly improve the calculated gaps, at a high computational cost. Here, we take a different, more flexible route, by mapping our DFT results on a simple model which is easier to use in calculations of experimental responses, and which can be adapted to compensate for the shortcoming of DFT calculations that is the value of the gap.

The double point group associated to the space group $Cmcm$ at $\Gamma$ is $mmm$. We calculated that the bands just below (respectively just above) the Fermi level are associated with the irreducible representations $\Gamma_5^+$ (respectively $\Gamma_5^-$) of this double point group. The character table for these two irreducible representations is:
\begin{center}
\begin{tabular}{c|cccccccccc}
 & $E$ & $-E$ & $C_2$ & $C_2'$ & $C_2''$ & $i$ & $-i$ & $s_v$ & $s_v'$ & $s_v''$\\
\hline
$\Gamma_5^+$ & 2 & -2 & 0 & 0 & 0 &  2 & -2 & 0 & 0 & 0\\
$\Gamma_5^-$ & 2 & -2 & 0 & 0 & 0 & -2 &  2 & 0 & 0 & 0
\end{tabular}
\end{center}

\begin{figure*}
\centering
\includegraphics[width=16cm]{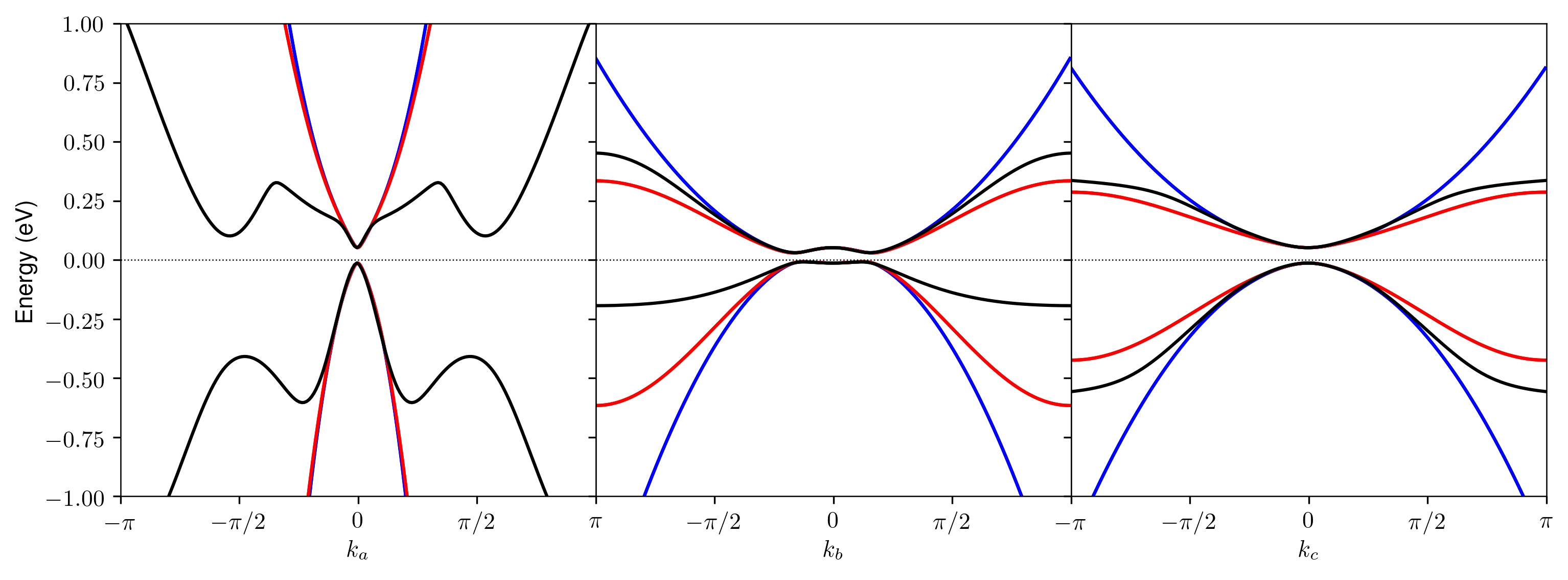}
\caption{Dispersion of the parabolic (blue) and lattice (red) models, compared to the two bands closest to the Fermi level in ZrTe$_5$ calculated using density functional theory (black). The black dotted line is the Fermi level as calculated by DFT.}
\label{fig:fit}
\end{figure*}

We constructed a model Hamiltonian for the four bands closest to the Fermi level using $\textbf{k} \cdot \textbf{p}$ theory, which yields Hamiltonians in powers of reciprocal coordinates constrained by symmetry. Since we are considering two irreducible representations, each of which has spin degeneracy, our Hamiltonian is four-dimensional. We chose $-E$, $C_{2y}$ and $m_{z}$ as a basis for $mmm$. The matrices for these operations in the basis of the spinor $\left( \left| \Gamma_5^+, \uparrow \right\rangle, \left| \Gamma_5^+, \downarrow \right\rangle, \left| \Gamma_5^-, \uparrow \right\rangle, \left| \Gamma_5^-, \downarrow \right\rangle \right)$ are:
\begin{align}
-E &=
\begin{pmatrix}
1 & 0 & 0 & 0\\
0 & 1 & 0 & 0\\
0 & 0 & -1 & 0\\
0 & 0 & 0 & -1
\end{pmatrix},
\end{align}
\begin{align}
C_{2y} &=
\begin{pmatrix}
0 & -i & 0 & 0\\
-i & 0 & 0 & 0\\
0 & 0 & 0 & -i\\
0 & 0 & -i & 0
\end{pmatrix},
\end{align}
\begin{align}
m_{z} &=
\begin{pmatrix}
0 & -1 & 0 & 0\\
1 & 0 & 0 & 0\\
0 & 0 & 0 & 1\\
0 & 0 & -1 & 0
\end{pmatrix}.
\end{align}
where $-E$ is the inversion operator in real space. To second order in $k_a$, $k_b$, and $k_c$, we obtain \cite{GreschThesis}:
\begin{widetext}
\begin{equation*}
H =
\begin{pmatrix}
C(\textbf{k})+M(\textbf{k}) & 0 & A_a k_a & A_b(k_a - 2k_b) -i A_c k_c \\
0 & C(\textbf{k})+M(\textbf{k}) & A_b(k_a - 2k_b) + i A_c k_c & -A_a k_a \\
A_a k_a & A_b(k_a - 2k_b) -i A_c k_c & C(\textbf{k})-M(\textbf{k}) & 0 \\
A_b(k_a - 2k_b) + i A_c k_c & - A_a k_a & 0 & C(\textbf{k})-M(\textbf{k})
\end{pmatrix}
\end{equation*}
\end{widetext}
where we used the functions:
\begin{align}
C (\textbf{k}) = C_0 - C_a k_a^2 - C_b (k_a k_b - k_b^2) - C_c k_c^2,\\
M (\textbf{k}) = M_0 - M_a k_a^2 - M_b (k_a k_b - k_b^2) - M_c k_c^2.
\end{align}
$k_a$, $k_b$ and $k_c$ are given in inverse lattice units. In the following, we refer to this as the parabolic model.

We determined the parameters of the model by fits to the band structure calculated with Elk close to $\Gamma$ along the three axes of the Brillouin zone. The fitted parameters are $A_a = 0.788$\ eV, $A_b = 0.021$\ eV, $A_c = -0.129$~eV, $M_0 = 0.033$\ eV, $M_a = -1.572$\ eV, $M_b = 0.127$\ eV, $M_c = -0.097$\ eV, $C_0 = 0.019$\ eV, $C_a = -0.569$\ eV, $C_b = 0.040$\ eV, and $C_c = -0.022$\ eV.

The quality of the fitted model is demonstrated in Fig.\ \ref{fig:fit}. Along $b$ and $c$, the effective mass is large and the fit holds for a larger portion of the Brillouin zone. However along $a$ the effective mass is much smaller, the DFT band structure has more features and the fit only holds for $|k_a| < 0.2$. Importantly, the model describes well the double-peak and double-well structure in the dispersion along $b$. The effective dispersion is very flat along this axis. This could be related to optical conductivity results which point to a parabolic dispersion along $b$ \cite{Martino2019}.

\begin{figure*}
\centering
\includegraphics[width=16cm]{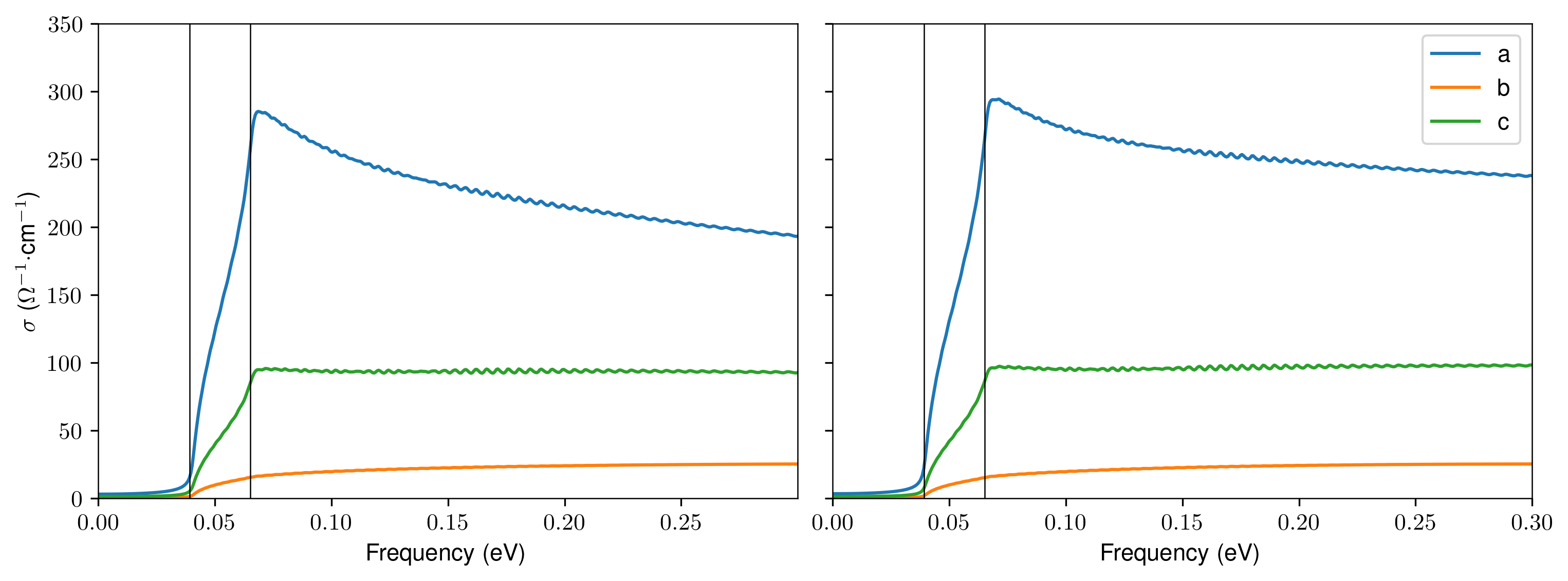}
\caption{Optical conductivity along $a$, $b$ and $c$ at $T=10$ K for the parabolic (left) and lattice (right) models. The vertical lines are the gap at $\Gamma$, 65 meV, and the minimum gap along $\Gamma-Z$, 41 meV.}
\label{fig:frequency-both}
\end{figure*}

We adapted the model to the case of a lattice by using the standard substitutions $k_i \rightarrow \sin(k_i)$ and $(k_i)^2 \rightarrow 2 \left[ 1-\cos(k_i) \right]$. The result is shown in Fig.\ \ref{fig:fit}, and compares well to the DFT band structure close to $\Gamma$. The widths of the lattice and DFT bands are close to each other along $b$ and $c$, while the DFT bands have a smaller width along $a$. The topological invariants of the lattice model are $(1,000)$, which are different from those of the DFT bands. This is unsurprising since the band inversion in ZrTe$_5$ happens away from $\Gamma$ \cite{Manzoni2016}, whereas the model parameters were obtained by fitting close to $\Gamma$; it is a common issue with $\textbf{k} \cdot \textbf{p}$ theory \cite{Nechaev2016, Choi2020}.

\section{Optical conductivity}

Now that we have justified the $\textbf{k} \cdot \textbf{p}$ theory model in comparison to \emph{ab-initio} calculations, we test its validity with respect to experimental data. We calculated the longitudinal optical conductivity along $a$ as \cite{TkachovBook}:
\begin{align}
\sigma_{a} (\omega,T) = &\frac{i e^2 a^2}{\hbar V} \sum_{n, n', \textbf{k}} \frac{|\left\langle n\textbf{k} \middle| \partial_{k_a} H \middle| n'\textbf{k} \right\rangle|^2}{\xi_{n'\textbf{k}} - \xi_{n\textbf{k}}}
\nonumber \\
&\times \frac{f(\xi_{n\textbf{k}},T) - f(\xi_{n'\textbf{k}}, T)}{\hbar\omega + \xi_{n\textbf{k}} - \xi_{n'\textbf{k}} + i\gamma}
\label{eq:Kubo}
\end{align}
where $\omega$ is the frequency, $T$ the temperature, $V$ the crystal volume, $n$ and $n'$ band indices, $\textbf{k}$ the crystal momentum, $\xi_{n \textbf{k}}$ denotes eigenvalues of $H$, $f$ is the Fermi-Dirac distribution, and $\gamma$ is a lifetime broadening from residual scattering.

The value of $\gamma$ can be related to the mean free path, which can be estimated as $v_F/\gamma$. The Fermi velocity along $a$ just above the Fermi level and off the $\Gamma$ point for $M_0$ between 3 and 33 meV is of the order of $4.5 \times 10^5$\ m/s. We choose $\gamma = 1$\ meV which yields a mean free path of 300 nm, in accordance with the experimental values $6.9 \times 10^5$\ m/s and 583 nm \cite{Martino2019, Yang2019}.

The calculated optical conductivity along $a$, $b$ and $c$ for the parabolic model as a function of frequency at $T=10$\ K is plotted in Fig.\ \ref{fig:frequency-both}. $\sigma_a (\omega)$ is vanishingly small at low frequencies, and rises sharply around 40\ meV, to reach around 300\ $\Omega^{-1}\cdot$cm$^{-1}$, which is very close to the value measured experimentally \cite{Martino2019}. There is a shoulder in this rise, which we attribute to the double well electronic structure along $b$. Also indicated are the values of the minimal gap along $b$ and of the gap at $\Gamma$, which coincide with the initial rise and levelling regime, respectively. At higher frequencies, the optical conductivity monotonically decreases. The optical conductivities along $b$ and $c$ follow a very similar trend. Their overall magnitudes differ by a factor three and both stay almost constant above the gap. The difference in magnitude between the optical conductivities along $a$ and $c$ is in line with the different flatnesses of the electronic bands along $k_a$ and $k_c$, and is close to the difference measured experimentally \cite{Martino2019}, which is also close to a factor of 3. The optical conductivity along $b$ is much smaller on account of the sizably reduced dispersion along $k_b$, and globally follows the evolution of the one along $c$, except that its rise at low frequency is smoother.

The optical conductivities for the parabolic and lattice models, plotted in Fig.\ \ref{fig:frequency-both}, are very similar to the results for the parabolic model as expected. The optical conductivity along $b$ has not been measured yet and therefore constitutes a prediction which could be used to confirm the validity of the model presented here.

The most important point in the evolution of the optical conductivity along $a$ and $c$ is that the high-frequency behaviour for $\hbar \omega > 60$ meV extrapolates to a finite value at zero frequency. This is a key signature that, according to Ref.\ \cite{Martino2019}, excludes the possibility of a Dirac dispersion close to $\Gamma$. Indeed, a Dirac dispersion would give a linear optical conductivity close to zero frequency. The experimental behaviour is successfully replicated by a two-band model with a parabolic dispersion along $b$, which yields an optical conductivity going as the square root of the frequency \cite{Martino2019}. In the same work, it is argued that this signature is also incompatible with a gapped Dirac model, which would give a linear optical conductivity with a shift in energy. At first sight, this is in contradiction with our results: the model we consider is a Dirac model with a \textbf{k}-dependent mass term. However the type of gapped Dirac model discussed in Ref.\ \cite{Martino2019} is very specific: it is a model where the dispersion is strictly linear above the gap, and features a discontinuous derivative at $\Gamma$. Our result therefore provides an alternative scenario in which the optical conductivities from above 60\ meV extrapolate to finite values at zero frequency. Moreover this scenario is in line with the proximity to a 3D Dirac state.

A second important point with respect to experiments is the presence of an inflection point in the sharp rise of the optical conductivity at low frequency. This matches the experimental data for $\sigma_a$ on one of the two samples in Ref.\ \cite{Martino2019}, modeled using a parabolic dispersion along $b$.

Finally, one point does not fit experiments: after the sharp rise, the slope of the optical conductivity along $a$ turns negative, unlike the experimental data where it keeps on rising but much slower \cite{Martino2019}. Our model is however entirely based on \emph{ab-initio} calculations, which are known not to give correct estimates of band gaps. In the next section, we use the flexibility of our model to modify the electronic structure by adjusting $M_0$, which in particular changes the gap to the experimental value.

\section{Adjusting the electronic structure}

Experimental measurements yielded a large range of band gaps: ARPES results alone range between 0 and 100\ meV \cite{Xiong2017}. Since we are considering a simple model, we can adjust it to fit various experimental gaps. The gap at $\Gamma$ in our model is $2 M_0$. We know that ARPES data fit DFT calculations away from $\Gamma$ \cite{Wu2016, Martino2019}, so we keep the other parameters fixed, and only modify $M_0$ to compare it specifically with the optical conductivity measured along $a$ in \cite{Martino2019}. The sample used for these measurements had a band gap of 6\ meV.

\begin{figure}
\centering
\includegraphics[width=8cm]{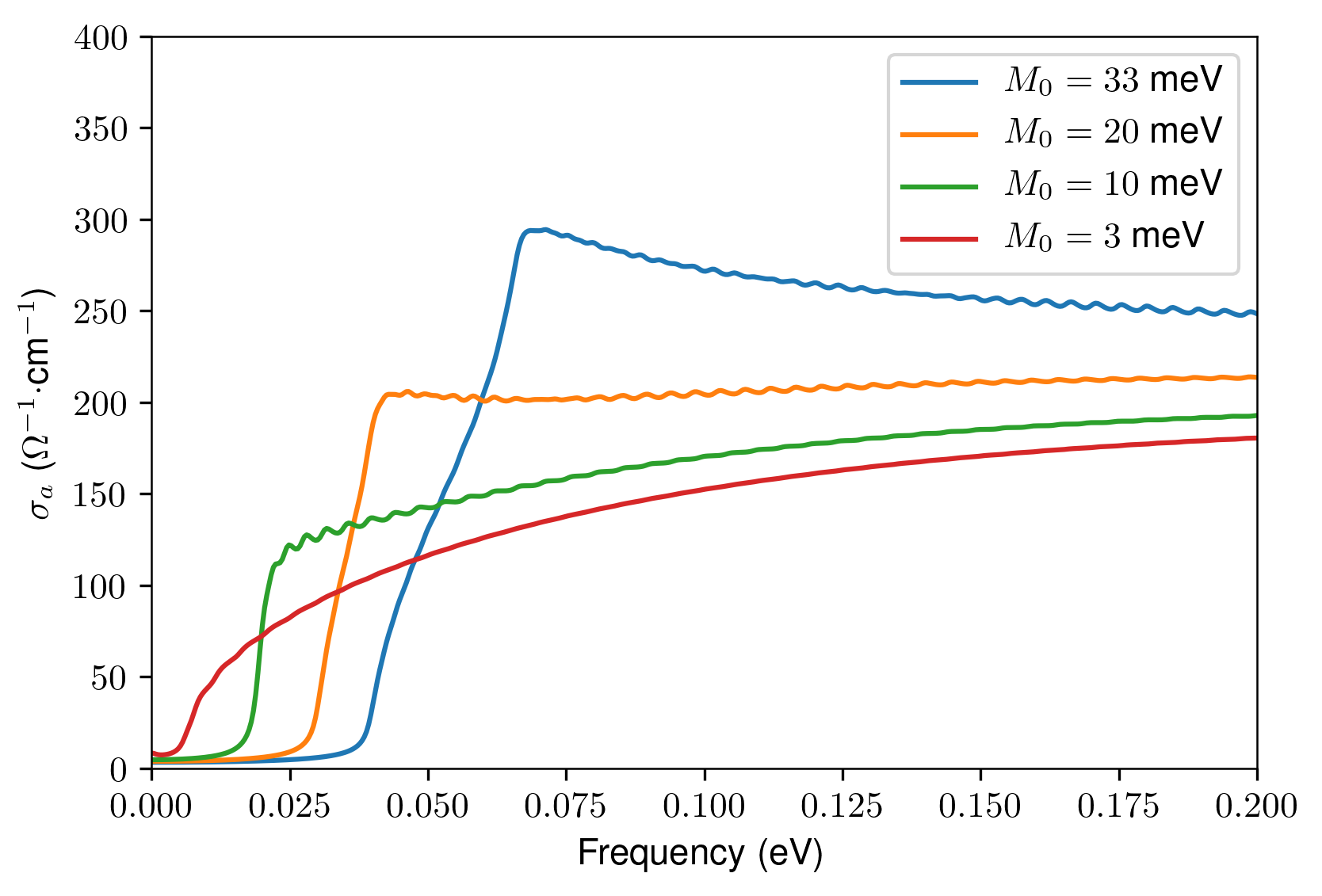}
\caption{Optical conductivity along $a$ at 10\ K for the lattice model, for four different values of $M_0$.}
\label{fig:frequency-gap}
\end{figure}

\begin{figure}
\centering
\includegraphics[width=8cm]{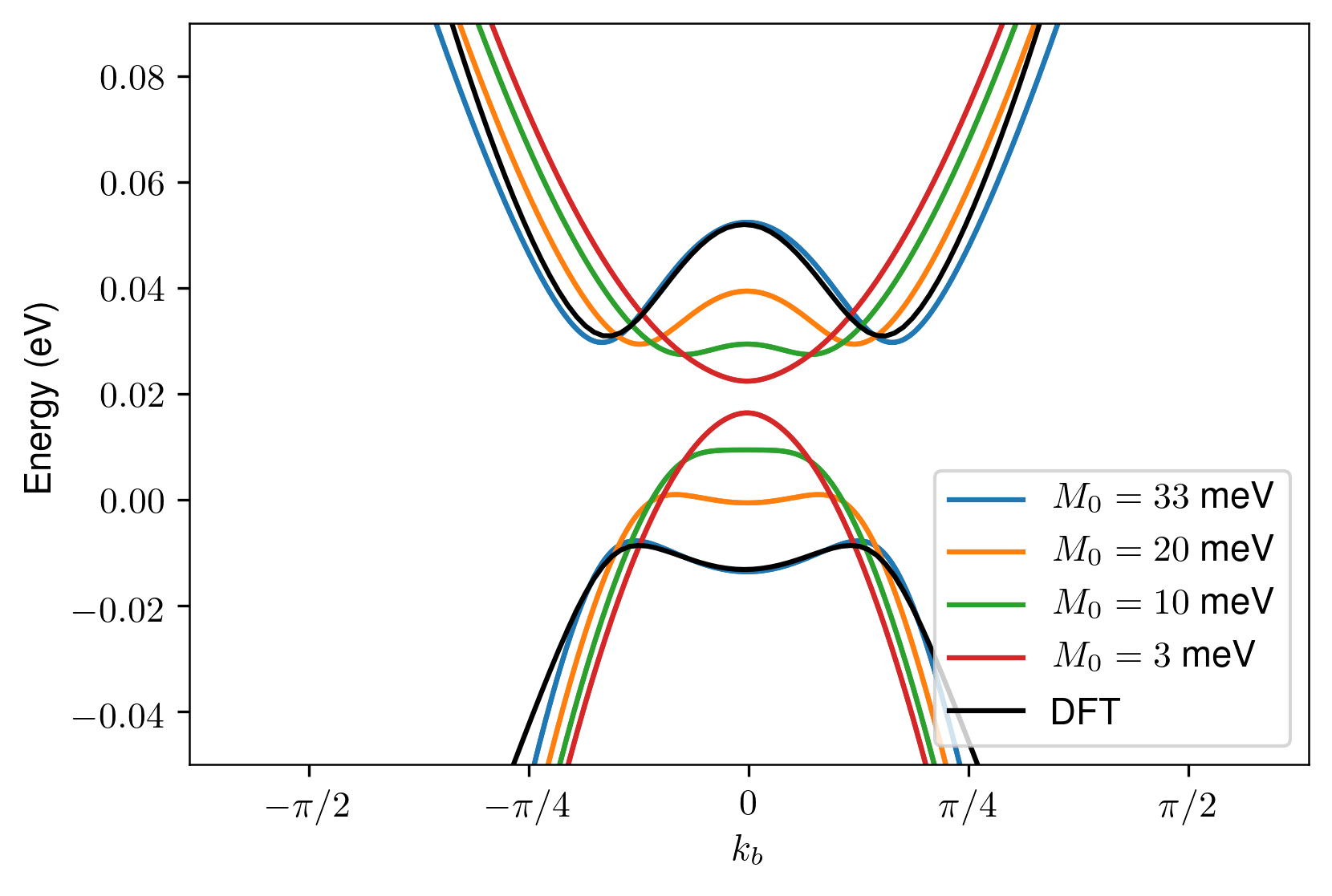}
\caption{Dispersion of the lattice model at $k_a=k_c=0$, for the four values of $M_0$ in Fig.\ \ref{fig:frequency-gap}. The double peak structure disappears when $M_0$ is lowered.}
\label{fig:spectrum-gap}
\end{figure}

We calculated the optical conductivity along $a$ for four values of $M_0$, ranging from 33 meV, which is the value given by DFT, to 3\ meV, which matches the experimental gap in \cite{Martino2019}. The results are displayed in Fig.\ \ref{fig:frequency-gap}. Lowering the gap has two main consequences: the overall magnitude is reduced, and the slope turns positive overall up to the peak close to $0.7$\ eV. Moreover, it still gives an extrapolation towards a finite value at zero frequency, and therefore fits overall very well the experimental data. It also comes very close to the behaviour of the two-band model introduced in \cite{Martino2019}, which goes as the square root of the frequency. The model we have put forward is thus a valid alternative to the one designed in Ref.\ \cite{Martino2019}, and bridges between \emph{ab-initio} calculations and experimental data.

Further comparison is possible with ARPES data which established that the gap opens at $\Gamma$ precisely, that is with $k_b=0$ \cite{Xiong2017}. This is unlike DFT results which, as discussed in Sec.\ \ref{Ab-initio calculations} locate the gap on the segment $\Gamma-Z$, i.e.\ away from $\Gamma$. This changes upon lowering $M_0$ to 3\ meV when the shoulders in the bands along $k_b$ disappear and the gap has moved to the $\Gamma$ point (Fig.\ \ref{fig:spectrum-gap}). This qualitative change of the shape of the spectrum can also be related to the qualitative change of shape of the optical conductivity upon changing $M_0$.

\section{Temperature dependence}

\begin{figure}
\centering
\includegraphics[width=8cm]{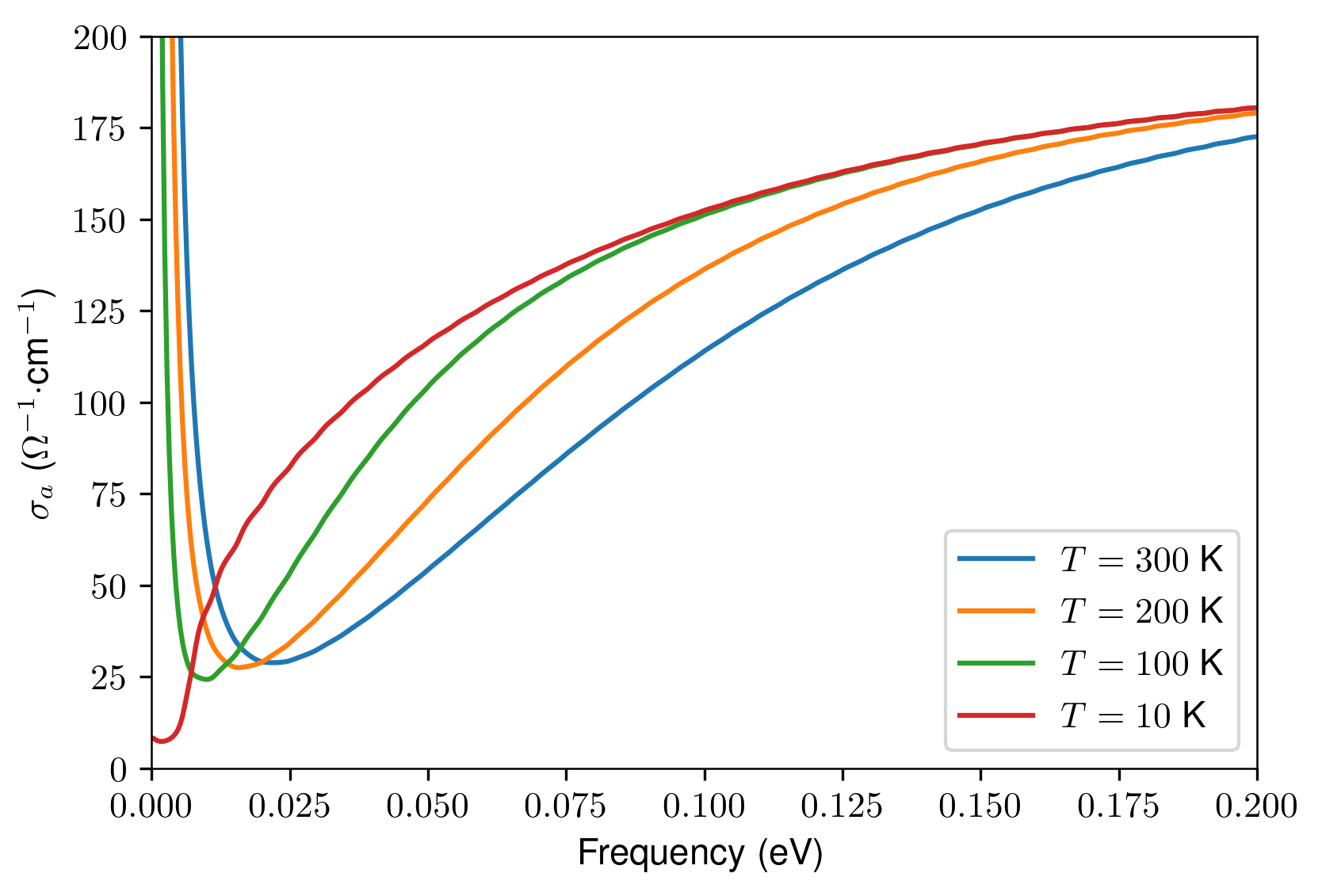}
\caption{Optical conductivity for the lattice model with $M_0=3$ meV, for four values of the temperature.}
\label{fig:frequency-temperature}
\end{figure}

\begin{figure}
\centering
\includegraphics[width=8cm]{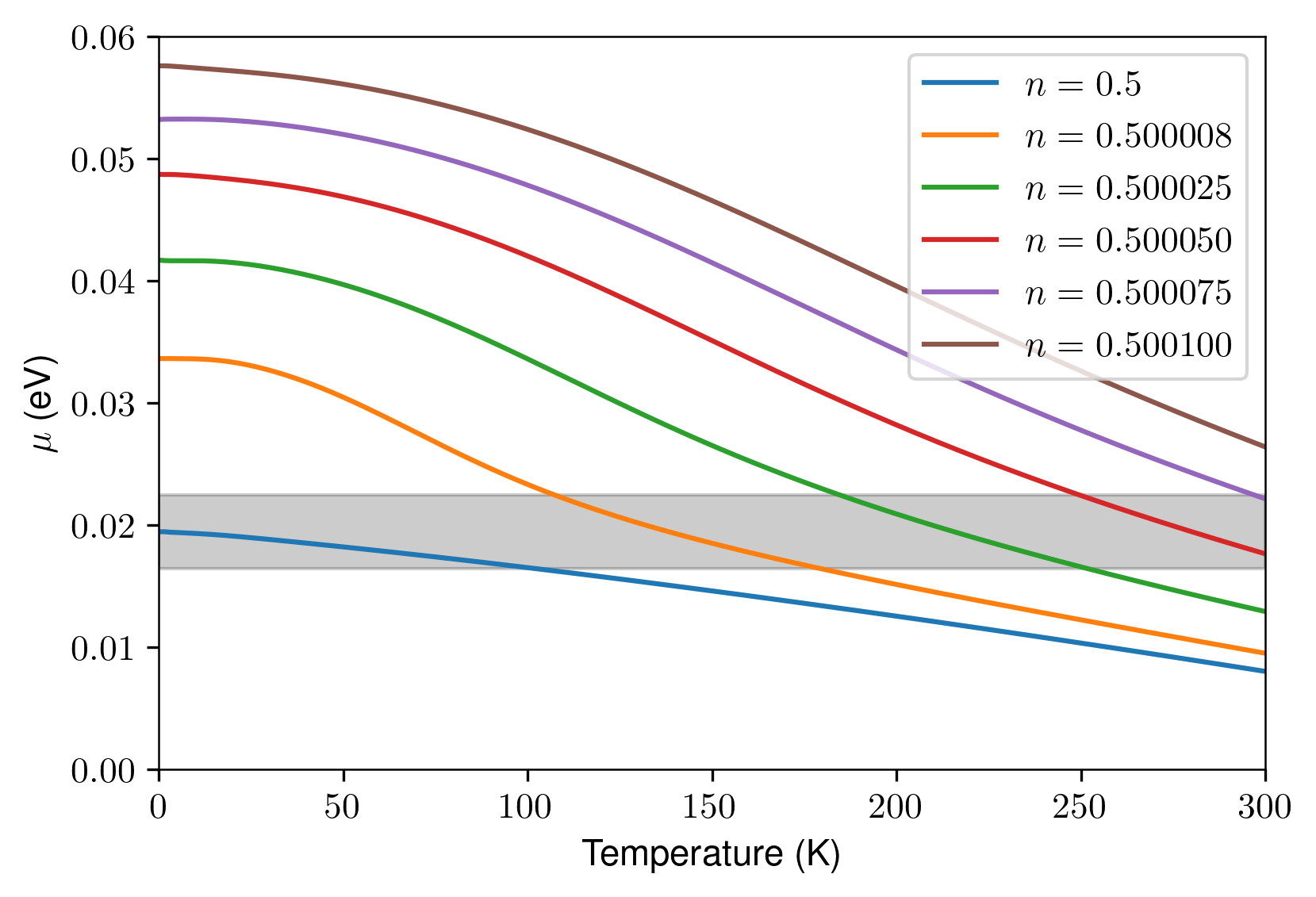}
\caption{Chemical potential as a function of temperature for the lattice model with $M_0=3$ meV, for various values of the filling. The grayed area represents the electronic gap.}
\label{fig:temperature-mu}
\end{figure}

In order to cover a larger range of experimental data, we evaluated the temperature dependence with $M_0=3$\ meV of the optical conductivity along $a$ (Fig.\ \ref{fig:frequency-temperature}). Strikingly, for temperatures down to 100\ K, $\sigma_a$ has a sharp Drude peak at low frequency. This is surprising since we are considering a semiconductor with a nominal Fermi energy inside the gap. To understand this better, we calculated the temperature evolution of the chemical potential $\mu$ (Fig.\ \ref{fig:temperature-mu}). At half filling, $\mu$ crosses the top of the valence band close to 100\ K, which explains the Drude peak above this temperature.

Moreover, we find that $\mu$ varies strongly with temperature, to the point that between 0 and 300\ K it sweeps an energy window larger than the gap. This matches ARPES measurements showing a large drop of $\mu$ when raising temperature \cite{McIlroy2004, Manzoni2015, Xiong2017, Zhang2017}, as well as NMR results \cite{Tian2019}. In addition, it is consistent with reports of strong variation of the carrier density with temperature in optical conductivity \cite{Chen2015} and Hall \cite{Tang2019} measurements. However, these experimental results are still being debated, since a raising of $\mu$ when raising temperature has also been measured using ARPES in some samples \cite{Moreschini2016, Wu2016}. The difference with the measurements showing a lowering was related to variations in sample growth conditions \cite{Xiong2017}.

This lowering of $\mu$ can also be seen as a change from $n$- to $p$-type carriers when raising the temperature. This was measured as a change of sign in the thermopower \cite{Jones1982, McIlroy2004, Shahi2018, Martino2019}, concomitant with a peak in the resistivity, and thus interpreted as its origin. For this reason, we calculate the conductivity and resistivity as a function of temperature.

So far we considered the four-band model at half filling; $\mu$ converges towards the middle of the gap at zero temperature. This implies that the system is insulating at low temperature and will have a diverging resistivity in the limit of zero temperature, unlike most experimental results. We know however from ARPES experiments that in many samples the chemical potential does not lie inside the gap \cite{McIlroy2004, Manzoni2015, Li2016a, Moreschini2016, Wu2016, Xiong2017, Zhang2017} which was explained by $n$-doping due to chemical vapor transport synthesis \cite{Shahi2018}.

\begin{figure*}
\centering
\includegraphics[width=16cm]{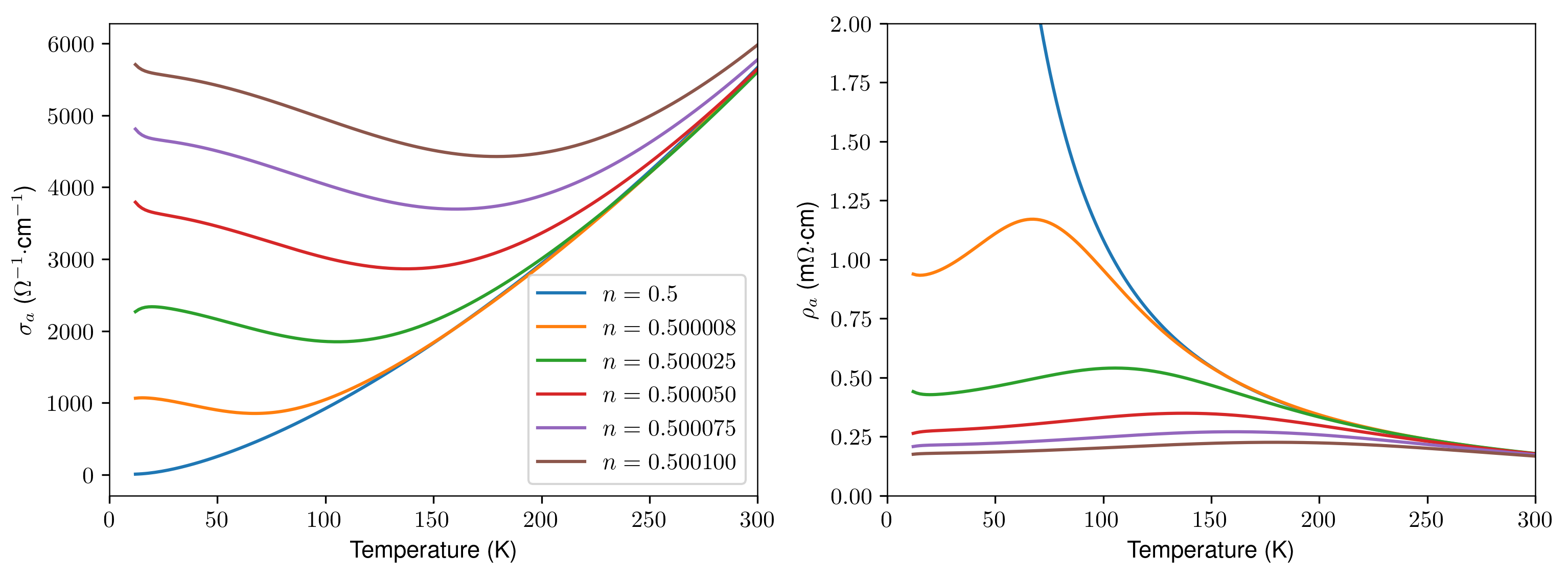}
\caption{dc conductivity $\sigma_a$ (left) and resistivity $\rho_a = \sigma_a^{-1}$ (right) as a function of temperature with $M_0=3$ meV, for various values of the filling.}
\label{fig:temperature-resistivity}
\end{figure*}

Our results for $\mu(T)$ for different values of the filling $n \geq 0.5$ are presented in Fig.\ \ref{fig:temperature-mu}. Already for a filling increase as small as from $0.5$ to $0.500008$, the chemical potential rises well above the gap towards zero temperature. Its change with temperature is also faster than in the half-filling case, it crosses the whole gap within less than 100\ K and lies well below the valence band edge at 300 K. When raising the filling above $0.500075$, $\mu$ is shifted upwards and no longer enters the gap below 300\ K and instead lies within the conduction band. This is particularly striking: ZrTe$_5$ undergoes a transition from semiconductor to metal for a variation in filling of the order of $10^{-4}$ ($=2.54 \times 10^{17}$ cm$^{-3}$), which is very small. This sheds light from a theoretical point of view on the strong variations in samples prepared with different synthesis methods \cite{Shahi2018}.

\begin{figure*}
\centering
\includegraphics[width=16cm]{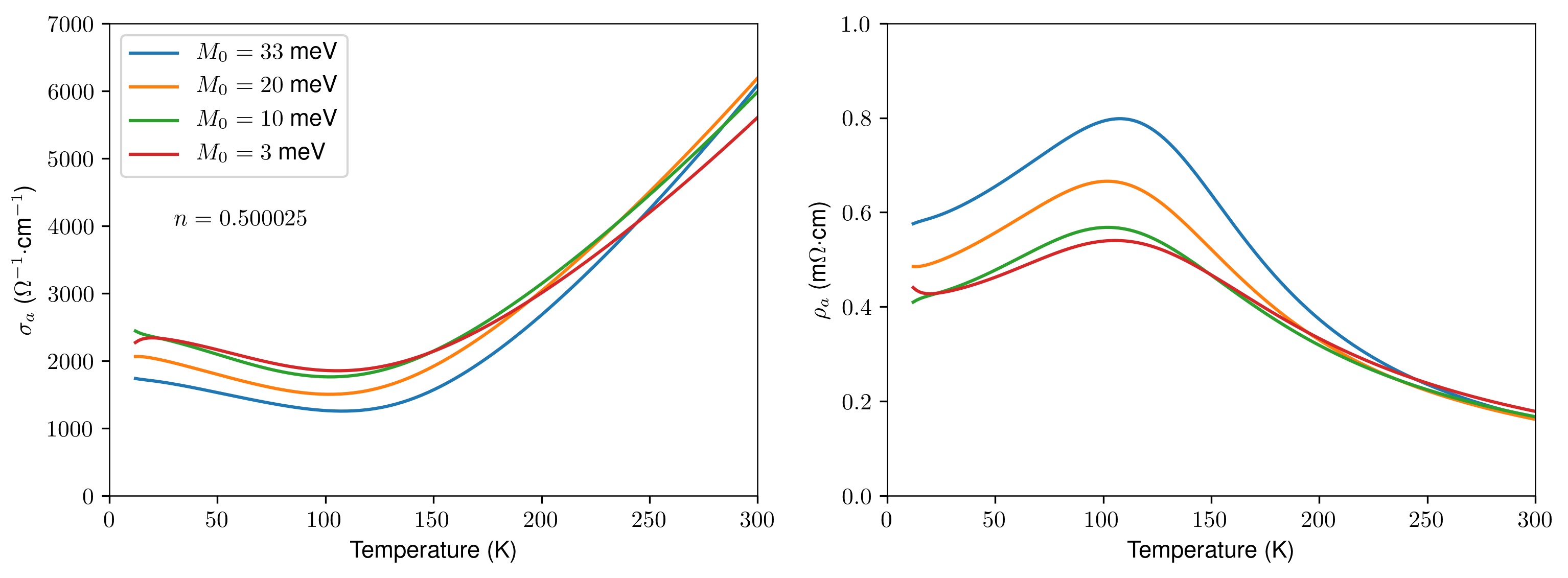}
\caption{Conductivity (left) and resistivity (right) at zero frequency as a function of temperature for the lattice model with a filling of $0.500025$ and four different values of $M_0$.}
\label{fig:temperature-resistivity-gap}
\end{figure*}

\begin{figure}
\centering
\includegraphics[width=8cm]{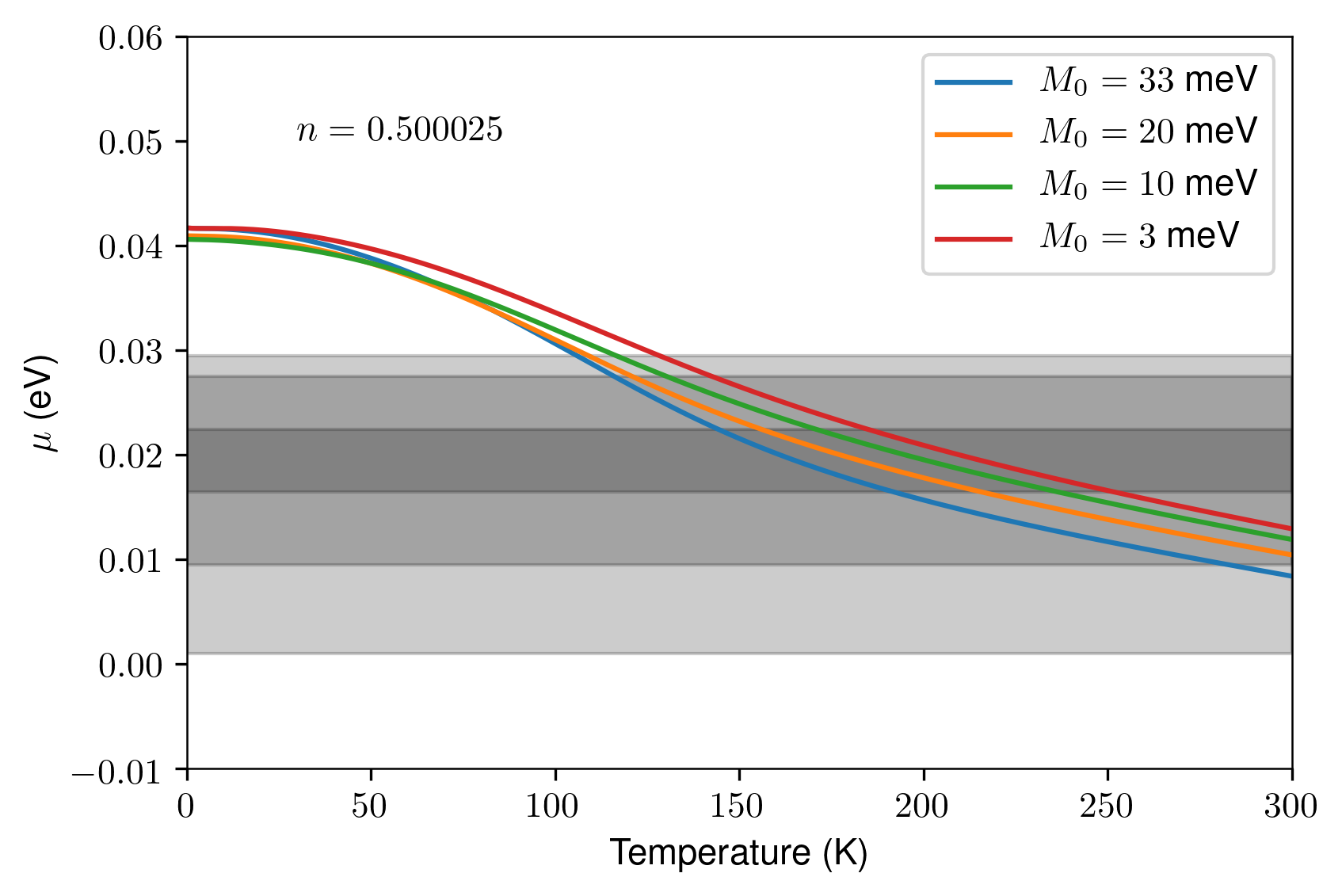}
\caption{Chemical potential as a function of temperature for the lattice model with a filling of $0.500025$ and four different values of $M_0$. The grayed areas represent the electronic gaps corresponding  to the three smallest values of $M_0$ considered.}
\label{fig:temperature-mu-gap}
\end{figure}

Results for the dc conductivity $\sigma_a$ and the resistivity $\rho_a = \sigma_a^{-1}$ as a function of temperature for different values of the filling are collected in Fig.\ \ref{fig:temperature-resistivity}. As expected, the resistivity at half filling diverges at zero temperature. When raising the filling, the resistivity at zero temperatures turns finite and a peak develops at intermediate temperatures. Its height goes down and its position shifts to higher temperatures when raising the filling, clearly showing that the crossing of the gap, and therefore the variation of $\mu$, is the cause of the resistivity peak.

The position of the resistivity peak is therefore a good comparison point between samples, as it is very sensitive to very small changes of filling. It has indeed been used as such in the literature, with samples in one study sometimes having resistivity peaks 50 K apart but with the same height \cite{Martino2019}.

We single out one filling value in order to study the dependence of the peak on the gap. We choose to focus on $0.500025$, at which $\mu$ crosses the 6\ meV gap entirely between approximately 190\ K and 250\ K. When the gap is raised, the temperature evolution of the chemical potential almost stays the same (Fig.\ \ref{fig:temperature-mu-gap}), but the physics changes: for $M_0 = 10$\ meV and above, the chemical potential does not reach the top of the valence band at 300\ K.

The calculated temperature-dependent resistivity, displayed in Fig.\ \ref{fig:temperature-resistivity-gap}, shows a prominent peak between 100 and 150\ K. The peak temperature does neither correspond to the temperature at which the chemical potential enters the gap, nor to the temperature where it reaches the middle of the gap, but rather falls in between the two. This is due to the rise in resistivity being compensated by the broadening due to higher temperatures. The peak height increases with the gap from 0.5\ m$\Omega\cdot$cm for $M_0=3$\ meV to 0.8\ m$\Omega\cdot$cm for $M_0=33$\ meV. Experimental results show resistivity peaks at temperatures ranging from 60\ K to 160\ K whose maximum varies between 0.8 and 3\ m$\Omega\cdot$cm \cite{Tritt1999, McIlroy2004, Chen2015, Manzoni2015, Li2016a, Zheng2016, Li2018, Xu2018, Tian2019, Tang2019, Martino2019} with which our results are consistent. Note that the peak height depends on $\gamma$: a larger value for $\gamma$ would decrease the height of the Drude peak and therefore increase the resistivity. The resistivity increases from 300\ K down to the peak temperature by a factor between 3 and 4 depending on the gap, as observed experimentally \cite{Martino2019}. It levels off towards zero temperature due to the temperature-independent residual scattering in the calculation of the conductivity. The zero-temperature limit increases both with decreasing filling and with increasing gap. The resistivity at very low temperature exhibits small down- or up-turns, related to the discrete \textbf{k}-point mesh used in the calculation, which becomes less fine than the sharp peaks in Eq.\ \ref{eq:Kubo} for low enough temperature.

\section{Conclusion}

We derived a four-band model for ZrTe$_5$ using $\textbf{k} \cdot \textbf{p}$ theory, and fitted its parameters on the DFT band structure. Since ARPES results match DFT results except for the value of the gap \cite{Wu2016, Martino2019}, we varied $M_0$ in the model to modify the electronic structure and in particular adjust the band gap to its experimental value, and left the other parameters untouched. The calculated optical conductivity corresponds well to experimental data \cite{Xu2018, Martino2019}, showing that the model, as well as DFT calculations, are instrumental in describing the physics of ZrTe$_5$ close to the Fermi level.

The chemical potential varies strongly with temperature, to the point that for fillings slightly above one half it crosses the gap entirely between zero and room temperature. The variation in filling needed to shift the chemical potential is very small, of the order of $10^{-4}$. This sheds light on how small variations in synthesis can cause large differences between samples with respect to their electronic conduction \cite{Shahi2018}. The temperature-dependent resistivity for various values of the filling displays a prominent peak; its position, width, and amplitude are consistent with experimental data. Moreover, its position shifts with the filling, and therefore the chemical potential. This confirms theoretically the conclusions of key experiments which attributed the origin of the resistivity peak to the large shift of the chemical potential with temperature \cite{McIlroy2004, Manzoni2015, Xiong2017, Zhang2017, Xu2018, Tian2019}. In particular, we obtain this resistivity peak without relying on a topological phase transition.

We have therefore, using a single model fitted on \emph{ab-initio} results, successfully replicated three key experimental results on ZrTe$_5$, namely the specific shape of the low frequency optical conductivity, the variation of the chemical potential with temperature and the peak in the temperature dependent resistivity. This renders this model a promising basis for addressing other exotic physics discovered in this material, such as its peculiar magnetotransport \cite{Liang2018}.

\section*{Acknowledgements}

We thank Daniel Braak, Liviu Chioncel, Andreas Östlin, Miloš M.\ Radonjić, Patrick Seiler, Alexey Shorikov and Yi Zhang for helpful discussions. This work was supported by the Deutsche Forschungsgemeinschaft --- Grant  number 107745057, TRR 80.

\bibliographystyle{apsrev4-1}
\bibliography{ZrTe5}

\end{document}